\documentclass[iop,revtex4]{emulateapj}
\usepackage{lineno}

\begin{document}

\renewcommand{\topfraction}{1.0}
\renewcommand{\bottomfraction}{1.0}
\renewcommand{\textfraction}{0.0}

\title{Speckle interferometry at SOAR in 2014\altaffilmark{\dag} }

\altaffiltext{\dag}{Based on observations obtained  at the Southern Astrophysical Research
(SOAR) telescope,  which is a  joint project of the  Minist\'{e}rio da
Ci\^{e}ncia,  Tecnologia, e  Inova\c{c}\~{a}o (MCTI)  da Rep\'{u}blica
Federativa do Brasil, the  U.S. National Optical Astronomy Observatory
(NOAO), the  University of  North Carolina at  Chapel Hill  (UNC), and
Michigan State University (MSU).}

\author{Andrei Tokovinin}
\affil{Cerro Tololo Inter-American Observatory, Casilla 603, La Serena, Chile}
\email{atokovinin@ctio.noao.edu}
\author{Brian D. Mason \& William I. Hartkopf}
\affil{U.S. Naval Observatory, 3450 Massachusetts Ave., Washington, DC, USA}
\email{bdm@usno.navy.mil, wih@usno.navy.mil}
\author{Rene A. Mendez}
\affil{Universidad de Chile,  Casilla 36-D, Santiago, Chile}
\email{rmendez@u.uchile.cl}
\author{Elliott P. Horch}
\affil{Department of Physics, Southern Connecticut State University, 501 Crescent Street, New Haven, CT 06515, USA}
\email{horche2@southernct.edu}

\begin{abstract}
The  results  of  speckle  interferometric observations  at  the  SOAR
telescope in  2014 are  given. A total  of 1641 observations  were taken,
yielding 1636 measurements of  1218 resolved binary and multiple stars
and 577 non-resolutions of 441 targets.  We resolved for the first
time  56 pairs,  including  some nearby  astrometric or  spectroscopic
binaries and  ten new subsystems in previously  known visual binaries.
The calibration of the data is checked by linear fits to the positions
of 41 wide  binaries observed at SOAR over  several seasons. The typical
calibration  accuracy is  0\fdg1 in  angle and  0.3\% in  pixel scale,
while the measurement errors are on the order of 3\,mas.  The new data
are  used here  to compute  194 binary-star  orbits, 148  of  which are
improvements  on  previous orbital  solutions  and  46 are  first-time
orbits.  
\end{abstract}

\section{Introduction}
\label{sec:intro}

Binary  stars   matter  in  astronomy  in  many   different  ways:  as
calibrators  of  various  stellar   properties,  as  tracers  of  star
formation,  and as hosts  to diverse  astrophysical phenomena  such as
mass  transfer,  circumstellar and  circumbinary  discs, or  dynamical
resonances, to name  a few. Knowledge of orbital  elements is needed in
these applications.   However, only a  small fraction of  known visual
binaries have  known orbits, mostly  because of orbital  periods being
much longer than  the time covered by observations.  Even known orbits
are not always reliable for the same reason: lack of sufficient data.

We report  here a  large set of  binary-star measurements made  at the
4.1-m  Southern  Astrophysical  Research  Telescope  (SOAR)  with  the
speckle camera. This paper  continues the previous series published by
\citet[][hereafter   TMH10]{TMH10},  \citet{SAM09},  \citet{Hrt2012a},
\citet{Tok2012a},  and \citet[][hereafter TMH14]{TMH14}.   Our primary
goal is improvement of known orbital elements and determination of new
orbits.  The emphasis  is placed on close and  nearby pairs resolvable
at SOAR, where orbital periods are measured in years rather than decades or
centuries.   The orbits  of these  fast  binaries can  be computed  or
improved only after a  few years of speckle-interferometry monitoring.
Many orbits were already computed for the first time or improved using
data obtained during previous speckle runs.   An ``orbit  optimizer'' was
used now to select  binaries where observations in the  coming three years
could potentially yield noticeable orbit improvement.

The second, overlapping program is the characterization of multiplicity of
solar-type stars within 67\,pc \citep[the FG-67 sample,][]{FG67a}.  By
resolving known binaries with astrometric acceleration \citep{MK05} or
variable radial  velocity (RV), we  obtain estimates of  their periods
and  mass ratios and  lay a foundation  for future  orbit determination.
The   spectroscopic   discoveries   are   mostly  furnished   by   the
Geneva-Copenhagen  Survey  of  \citet[][hereafter GCS]{N04}.   On  the
other hand, by observing  relatively wide binaries with separations up
to  3\arcsec ~we constrain  the existence  of subsystems  around their
components.   Several   such  subsystems  are   discovered  here.   In
addition, we  have now  specially targeted distant  physical secondary
components to the main FG-67 stars, looking for subsystems.  This work
\citep{Tok2014} complements the large  survey of the northern sky done
with Robo-AO  \citep{RAO} and shows  that subsystems in  the secondary
components are as frequent as in the main components.

In this paper we also observed {\it Hipparcos} binaries within 200\,pc
with   southern  declinations,  so   far  largely   neglected.   These
observations allow us to evaluate  orbital motion and to find a subset
of ``fast movers'' for  further monitoring. Some nearby tight binaries
from  {\it Hipparcos}  were  monitored  from the  outset  of the  SOAR
speckle  program.  This  strategy  of focusing  on  fast binaries  has
resulted in several new orbits, with more still to come.  Our approach
for this  program was therefore  similar and complementary to  that of
the speckle program active at the WIYN 3.5-m telescope at Kitt Peak in
recent  years  \citep[e.g.][]{Horch2011a,  Horch2011b,Horch2012}.   To
date, that  program has  resolved over 150  close companions  from the
{\it Hipparcos} list.

We  re-observed some  close (hopefully,  fast) pairs  resolved  at the
Blanco telescope in 2008. Some of them moved substantially in 6 years,
and may yield  first orbits. As a ``filler'',  several neglected pairs
were also measured or found unresolved.

This paper  is structured as  follows. Section~\ref{sec:obs} describes
briefly the  instrument and data  processing and gives an  estimate of
data consistency  for the  whole series of  measurements at  SOAR. The
results  are presented in  Section~\ref{sec:res} in  the form  of data
tables  with  comments  on  the  newly  resolved  pairs  and  multiple
systems. In Section~\ref{sec:orb}, the new and revised orbits are
given, and Section~\ref{sec:concl} concludes the paper.

\section{Observations}
\label{sec:obs}

\subsection{Instrument and observing method}

The   observations  reported   here  were   obtained  with   the  {\it
high-resolution camera} (HRCam) -- a fast imager designed to work at
the  4.1-m SOAR  telescope  \citep{TC08}. For  practical reasons,  the
camera was  mounted on  the SOAR Adaptive  Module \citep[SAM,][]{SAM}.
However,  the laser guide  star of  SAM was  not used;  the deformable
mirror   of  SAM   was  passively   flattened  and   the   images  are
seeing-limited. The SAM module corrects for atmospheric dispersion
and helps  to calibrate the pixel  scale and orientation  of HRCam, as
explained  in TMH14.   The transmission  curves of  HRCam  filters are
given in the instrument manual.\footnote{
\url{http://www.ctio.noao.edu/soar/sites/default/files/SAM/\-archive/hrcaminst.pdf}}
We  used  mostly  the  Str\"omgren  $y$ filter  (543/22\,nm)  and  the
near-infrared  $I$ filter  (788/132\,nm).  The  response curve  of the
latter  was re-defined  as a  product of  the filter  transmission and
detector response furnished by the respective manufacturers.

The electron multiplication CCD  (EM CCD) Luca-S used in HRCam failed
in 2014 July.  It was sent to its manufacturer  (Andor) for repair and
received back in 2014 December. For  the two runs in the fall of 2014,
we used the EM CCD Luca-R,  kindly loaned by G.~Cecil. It has a larger
format of  1004$\times$1002 pixels,  with smaller pixels  of 8\,$\mu$m
compared  to Luca-S (658$\times$496,  10\,$\mu$m pixels).  The second
lens in  HRCam was replaced by  the achromatic doublet  of 75-mm focal
length to  get the  pixel scale of  14.33\,mas. The  detector software
driver was also upgraded on this occasion. 

Although the two Luca cameras look  exactly the same and come from the
same manufacturer, their CCDs  are radically different. Luca-S uses a
line-transfer CCD,  while Luca-R  has a frame-transfer CCD.   We found
that Luca-R had poor charge  transfer efficiency (CTE) in the vertical
direction. Weak signals such as cosmic-ray hits and dark  current from
hot pixels are  smeared vertically over $\sim$ 5  pixels, while strong
signals  are  transferred  with  less  spread.  This  results  in  the
signal-dependent loss  of resolution in the  vertical direction (along
CCD  columns), which  was  accounted  for in  the  data processing  by
an additional  adjustable parameter.   This   problem   was   revealed
unexpectedly during the  2014 October run. In the  next run, we placed
stars closer to the line register, improving slightly the vertical CTE
by reducing the number of charge transfers.

We studied the distribution of the bias signal of both cameras.  It is
very well modeled  by the sum of a  Gaussian component (readout noise)
and a negative exponent that corresponds to the single-electron events
amplified  stochastically  by  the  gain register.   Most  events  are
generated by the CCD  clocking (clock-induced charge, CIC) typical for
EM CCDs.  The width (decrement) of the exponential term depends on the
EM gain.  The  rate of CIC events  is found as the ratio  of the areas
below  these  two  terms.   For  Luca-R,  the  readout  noise  is  4.5
Analog-to-Digital Units (ADU) and the CIC rate is 0.03. The CIC events
are not smeared vertically because no charge transfer is involved. For
Luca-S, the readout  noise is 9 ADU, the  exponential decrement is 30
ADU  (for the gain  setting of  200 used  here), and  the rate  of CIC
events is 0.13 per pixel. About  half of the CIC events are removed by
the threshold of 17~ADU applied in the power-spectrum calculation.

The  SOAR telescope  suffers from  50-Hz vibration  (see  TMH14).  The
vibrations are non-stationary, causing  variable loss of resolution if
the  20-ms  exposure time  is  used. Most  data  were  taken with  an
exposure  of 5\,ms,  sampling 1/4  of the  vibration period.  The
resolution  is then recovered  and the  effect of  vibrations  becomes less
dramatic.   It disappears completely  at an exposure  time of  2\,ms, 
used on the brightest targets. With such short exposures, the
power   spectra  extend  to   the  cutoff   frequency  and   are  very
symmetric. The Luca-R detector suffering from the CTE problem was used
mostly with  a 10-ms exposure  as a compromise between  vibrations and
poor CTE  at low signal levels.

The  observations  consist   of  taking  a  cube  of   400  images  of
200$\times$200  pixels each.   For binaries  wider than  1\farcs5, the
400$\times$400 format was used.  Each object and filter combination is
normally recorded  twice, these data are  processed independently, and
the result  is averaged.  We  used 2$\times$2 detector binning  on the
faintest targets observed in the $I$ filter, with a minor under-sampling
and  loss  of resolution.   Measurements  of  binaries  made with  and
without binning  agree well  mutually.  Faint  red dwarfs  with $I
\sim 12$ still produced useful data.

\subsection{Observing runs}

The  observing  time  for  this  program was  allocated  through  NOAO
(proposals 2013B-0172, 2014A-0038, 2014B-0019,  8 nights total) and by
the Chilean National TAC (proposal CN2014B-27, 4 nights).

\begin{deluxetable}{ l l r r r } 
\tabletypesize{\scriptsize}    
\tablecaption{Observing runs
\label{tab:runs} }                    
\tablewidth{0pt}     
\tablehead{ \colhead{Run}  &
\colhead{Dates}  & 
\colhead{$\theta_0$} & 
\colhead{Pixel} &
\colhead{$N_{\rm obj}$} \\
 &   &  \colhead{ (deg)} & \colhead{ (mas) } & 
}
\startdata
1 & 2014 Jan 15-16,22& 1.26     & 15.23 & 309  \\ 
2 & 2014 Mar 7-8     & $-1.30$  & 15.23 &  201  \\ 
3 & 2014 Apr 19-20   & $-1.40$  & 15.23 & 385  \\ 
4 & 2014 Oct 5-7     &  $-0.69$ & 14.33 & 553  \\
5 & 2014 Nov 7-8     &  0.66   & 14.33 &  253  \\
6 & 2015 Jan 10-11   & $-$0.20  & 15.23 &  248 \\ 
\enddata
\end{deluxetable}

Table~\ref{tab:runs}  lists   the  observing  runs,   the  calibration
parameters (position angle offset  $\theta_0$ and pixel scale in mas),
and  the number of  objects covered  in each  run. The  calibration of
angle  and  scale  was done  by  referencing  to  the SAM  imager,  as
described in TMH14.

Run 1  was affected by transparent  clouds.  Two hours on  2014 Jan 22
were added as a backup  program.  During this run, the Nasmyth rotator
of SOAR had a large offset of +2.8$^\circ$. The seeing was mediocre to
poor.  The instrument was not dismounted between runs 1 and 2, but the
Nasmyth  rotator  was  re-initialized,  explaining the  difference  in
$\theta_0$.  The sky was clear, and two hours were added on March 7 as
a backup from other program. The  sky was also clear on the two nights
of  run 3, with seeing good to excellent.

Runs 4 and 5 used the substitute detector Luca-R and were affected
by the CTE  problem.  For this reason the  observations were performed
mostly in the $I$ filter with  10-ms exposure. All three nights of 
run 4 were clear, and mostly with  good seeing (the median full width
at half  maximum, FWHM, of  re-centered average images  was 0\farcs60,
the best FWHM was around 0\farcs45).  The detector was removed between
 runs  4 and  5  in  an attempt to  improve  the  CTE  by tuning  the
electronic parameters; however, this turned  out to be impossible.  During 
run 5, the seeing varied from  average to poor. In the worst case, the
star did not  fit in the 3\arcsec ~field,  illuminating all pixels.  In
such cases, the image is truncated, the power spectrum contains bright
vertical  and horizontal  lines, and  the detection  limits  are poor.
Under poor  seeing,  data  obtained with the 400-pixel field  and 2$\times$2
binning are of better quality, without image truncation.

The two nights  of  run 6 were clear, with seeing  average to poor. The
HRCAM  was returned  to its  original configuration  with  the Luca-S
detector. Measurements  with the  SAM internal light  source confirmed
that the pixel scale did not change compared to  runs 1--3.

\subsection{Data processing}

The data processing is described in TMH10. As a first step, power
spectra and average re-centered images are calculated from the data
cubes. The auto-correlation functions (ACFs) are computed from the
power spectra. They are used to detect companions and to evaluate the
detection limits. The parameters of binary and triple stars are
determined by fitting the power spectrum to its model, which is a
product of the binary (or triple) star spectrum and the reference
spectrum. We used as a reference the azimuthally-averaged spectrum of
the target itself in the case of binaries wider than 0\farcs1. For
closer pairs, the ``synthetic'' reference was used (see TMH10). 

The  signal-dependent vertical smearing of  the  image 
caused  by the  CTE  problem required  a modification  of the  reference
spectrum.  We model the smearing by an additional Gaussian term in the
reference spectrum $P(f_x, f_y)$:
\begin{eqnarray}
P(f_x, f_y) & = &  T_{\rm DL}(\kappa)
\; 10^{ p_0 + p_1 \kappa} \; e^{- 2\pi^2
  (\Delta y f_y /2.35)^2 } , \\
\kappa  & = &  \sqrt{f_x^2 + f_y^2}/f_c , \nonumber
\label{eq:ell}
\end{eqnarray}
where  $f_x$ and  $f_y$ are  spatial frequency  coordinates  along CCD
lines  and  columns  respectively,  $T_{\rm DL}$ is  the  diffraction-limited
transfer  function, $f_c$  is the  cutoff frequency,  $\kappa =  |f| /
f_c$, $p_0$  and $p_1$ describe  the speckle signal  attenuation (both
are  negative, $p_0$  is related  to  seeing), and  the new  parameter
$\Delta y$  is the FWHM  of the CTE  smear in the column  direction in
pixels.   This ``elliptical''  synthetic  reference was  used for  all
binaries,  both  wide  and  close,  observed  in  runs  4  and  5.  The
parameters of the reference $(p_0,  p_1, \Delta y)$ were fitted to the
power  spectrum jointly  with  the binary  parameters $(\theta,  \rho,
\Delta m)$.

In  the case  of  close  binaries, the  parameters  of the  elliptical
reference and  of the  binary itself are  mutually correlated,  so the
resulting measures could be biased.  The close binary star B~430 ({\it
  19155$-$2515}) was  observed with position angles  of the instrument
differing by  $90^\circ$.  The ``fringes''  in the power  spectrum had
different  orientation  relative  to  the  CCD.  However,  the  binary
parameters $(\theta,  \rho, \Delta m)$  fitted to these  power spectra
are mutually consistent, (283\fdg7, 0\farcs0673, 0.785) and (285\fdg0,
0\farcs0690,  0.722). Therefore, the  CTE effect  is, to  first order,
accounted for by the modified reference model.  Still, measurements of
close binaries from   runs 4 and 5 (2014.77  and 2014.86) should be
taken with some caution.

Wide binaries  resolvable in the re-centered  long-exposure images are
processed  by another  code that  fits only  the  magnitude difference
$\Delta m$,  using the binary  position from speckle  processing. This
procedure  corrects   the  bias  on  $\Delta  m$   caused  by  speckle
anisoplanatism  and establishes the  correct quadrant  (flag *  in the
data table).

For  run  5, we  also calculated   shift-and-add (SAA  or ``lucky'')
images, centered  on the  brightest pixel in  each frame  and weighted
proportionally  to the intensity  of that  pixel.  All  frames without
rejection were  co-added.  Binary  companions, except the  closest and
the  faintest ones,  are detectable  in these  SAA images,  helping to
identify the correct quadrant. Such cases  are marked by the flag q in
the data table.   Quadrants of the remaining binary  stars are guessed
based on prior data or orbits,  not measured directly.

\subsection{Recalibration}

\begin{figure*}
\plotone{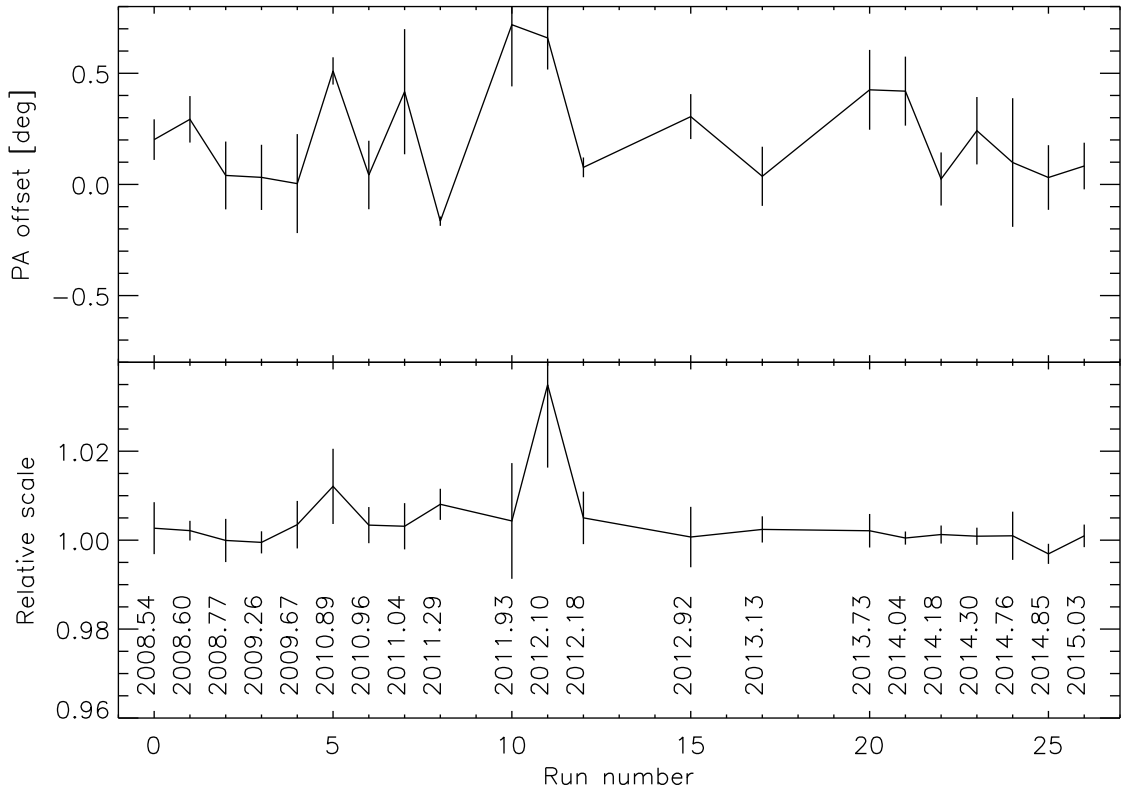}
\caption{\label{fig:cal} Systematic offsets  in angle (in degrees) and
  relative scale determined from wide binaries. The beginning date of each
  run is shown in the lower plot. }
\end{figure*}

Calibration of  scale and angular offset in  speckle interferometry is
challenging  because  the accuracy  of  modern  measurements with  4-m
telescopes exceeds  the accuracy of even the  best orbits.  Comparison
with ephemerides  is useful  only as a  sanity check.   Recognizing this
problem, we observed several wide  binaries  during each run. Their motions
are slow and  can be modeled by linear functions  of time. These models
can then be used to check  the calibration of the archival data.  This
approach tests only the mutual consistency of  speckle runs, rather
than absolute calibration of the whole data set.

We  selected 41  binaries  wider  than 0\farcs5  that  do not  contain
resolved subsystems  and were  observed during at  least 4 runs  each. The
angle  and  separation  of  each  binary  is  approximated  by  linear
functions of  time. If  the fitted  slope is less  than twice  its rms
error,  a zero  slope is  assumed.  Deviations  from these  models are
interpreted  as  calibration  corrections.   For each  run,  they  are
median-averaged. We then  iterate  by  fitting  new  linear  models
corrected  for the  run's  systematics, determining  new  calibration
parameters of the runs, etc.

Figure~\ref{fig:cal} shows  the offsets in angle  and scale determined
by  this procedure  for those  of  26 runs  where at least   $N=2$
calibrators were observed.  Run 0 refers to  observations at the
Blanco  telescope in  2008.5 (TMH10).   Runs 1--6  in this  paper
correspond to numbers  21--26 in the plot. Vertical bars show the
rms  scatter between  calibrators (not  errors of  the mean,  which are
smaller by $\sqrt{N}$). A typical rms scatter is 0\fdg1 in angle and
0.3\% in scale.

This analysis  reveals good  mutual consistency of  $\sim$5000 speckle
measurements produced  by HRCam to date.  The  largest systematics are
found in runs 10 and 11 (2011.9 and 2012.1). The published data can be
corrected by  subtracting the angular offsets found  here and dividing
the  separations by the  new scale  factors. We  plan to  observe more
calibrators  and  publish   improved  systematic  corrections  in  the
future. The  data of this paper  rely only on  the original instrument
calibrations and {\em are not corrected} for the offsets found so far.

This study gives an estimate  of the accuracy of speckle measurements.
After correction  for run systematics, the median  rms deviations from
the  models for the  41 calibrator  binaries are  0\fdg1 in  angle and
1.9\,mas  in separation.  The  actual  errors should  be  a factor  of
$\sim$1.5  larger  ($\sim$3\,mas),  considering  that  the  number  of
independent  measurements   is  about  twice  the   number  of  fitted
parameters. This study  will be repeated in the  future with additional
speckle runs and calibrators.

\section{Results}
\label{sec:res}

\subsection{Data tables}

The data tables have almost the  same format as in the previous papers
of  this  series. They  are  available  in  full only  electronically.
Table~2  lists  1636  measures  of  1218  resolved  binary  stars  and
subsystems, including 56 newly  resolved pairs. The columns of Table~2
contain  (1) the  WDS  \citep{WDS} designation,  (2) the  ``discoverer
designation'' as adopted  in the WDS, (3) an  alternative name, mostly
from the {\it Hipparcos}  catalog, (4) Besselian epoch of observation,
(5)  filter,  (6) number  of  averaged  individual  data cubes,  (7,8)
position angle  $\theta$ in degrees and internal  measurement error in
tangential direction $\rho  \sigma_{\theta}$ in mas, (9,10) separation
$\rho$ in  arcseconds and its  internal error $\sigma_{\rho}$  in mas,
and  (11) magnitude  difference  $\Delta m$.  An  asterisk follows  if
$\Delta  m$ and  the true  quadrant are  determined from  the resolved
long-exposure image;  a colon  indicates that the  data are  noisy and
$\Delta m$ is likely over-estimated  (see TMH10 for details); the flag
``q" means the  quadrant is determined from the  SAA image.  Note that
in the cases of multiple  stars, the positions and photometry refer to
the  pairings  between  individual  stars, not  the  photo-centers  of
subsystems.

For  stars with known  orbital elements,  columns (12--14)  of Table~2
list the residuals to the  ephemeris position and code of reference to
the         orbit         adopted         in   the Sixth Catalog
\citep[][hereafter VB6]{VB6}.\footnote{See
  \url{http://ad.usno.navy.mil/wds/orb6/wdsref.html}}  New and revised
orbits computed in this work (Section~\ref{sec:orb}) are referenced as
``This work''.

We did not  use image reconstruction and measured  the position angles
modulo  $180^\circ$  (except the  SAA  images  in  run 5).   Plausible
quadrants are assigned  on the basis of orbits  or prior observations,
but they can  be changed if required by  orbit calculation. For triple
stars, however, {\em both} quadrants  of inner and outer binaries have
to  be changed  simultaneously; usually  the slowly-moving  outer pair
defines the quadrant of the inner subsystem without ambiguity.

Table~3 contains the  data on 441 unresolved stars,  some of which are
listed   as  binaries   in  the   WDS  or   resolved  here   in  other
filters. Columns (1) through (6)  are the same as in Table~2, although
Column (2)  also includes other  names for objects  without discoverer
designations.   For  stars  that  do  not have  entries  in  the  WDS,
fictitious WDS-style codes  based on the position are  listed in Column
(1).  Column (8) is the estimated resolution limit, the largest of the
diffraction radius  $\lambda/D$ and the vertical CTE  smear $\Delta y$
(applicable to runs 4 and 5  only).  Columns (8,9) give the $5 \sigma$
detection  limits $\Delta  m_5$ at  $0\farcs15$ and  $1''$ separations
determined by the procedure described  in TMH10 (please note that this
is not  the resolution  of the observations).   When two or  more data
cubes  are processed,  the largest  $\Delta  m$ value is  listed. The  last
column marks by colons noisy  data mostly associated with faint stars.
In such cases, the quoted  $\Delta m$ might be too large (optimistic);
however,  the  information  that  these  stars were  observed  and  no
companions were found is still useful for statistics \citep{Tok2014}.

\subsection{Newly resolved pairs}
\label{sec:new}

\begin{figure*}[ht]
\epsscale{1.1}
\plotone{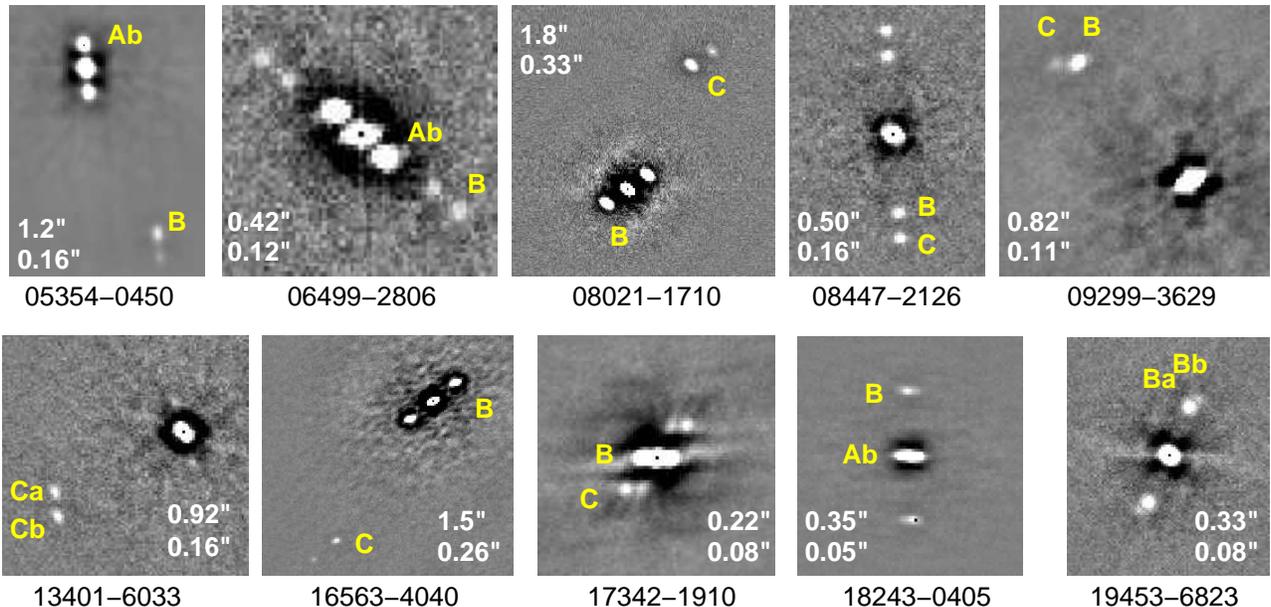}
\caption{\label{fig:ACF}  Fragments of ACFs  of newly  resolved triple
  systems (North  up, East  left, arbitrary scale).  The peaks  in the
  ACFs are labeled by component designations. Angular separations in
  the wide and close pairs are listed in the images. }
\end{figure*}

Table~4  lists  56 newly  resolved  pairs.   The last  two
columns of Table~4 contain  the spectral type (as given in
SIMBAD or  estimated from absolute magnitude) and  the {\it Hipparcos}
parallax \citep[][hereafter  HIP2]{HIP2}.  Fragments of  ACFs of newly
resolved triple systems are  shown in Figure~\ref{fig:ACF}. We comment
on the newly resolved  binaries below. The following abbreviations are
used: PM --  proper motion, CPM -- common proper  motion, RV -- radial
velocity,  SB1  and  SB2  -- single-  and  double-lined  spectroscopic
binaries, INT4 - 4th Interferometric Catalog \citep{INT4}.
\footnote{ {\em Note added in proof.} The  component C  in  16563$-$4040 (TOK  412 AC)  was independently
discovered by Sana et al. (2014, ApJS, 215, 15) and is designated in the
WDS  as  SNA 60 AC.  By error, the  newly  discovered  close companion  in 
04308$-$5727  (TOK 429 Aa,Ab)  was omitted  from Table  4, Figure  2 and
Section 3.2, but its measurements are found in Table 2.}

{\it 01379$-$8259.}  HIP~7601 is a  nearby (27\,pc) dwarf also known as
GJ~67.1  or HR~512.  According  to \cite{Wichman2003},  it is  a young
spectroscopic  triple detected  in  X-rays (1RXS  J013755.4$-$825838).
GCS  also  recognized  the  star  as  SB2.   Spectroscopic  monitoring
(Tokovinin,  in  preparation)  shows  that all  three  components  are
similar stars of  approximately one solar mass.  Here  we resolved the
outer  subsystem AB  and observed  its fast  motion.  In  fact  it was
already resolved at  SOAR on 2011.036 at 221\fdg0  and 0\farcs044, but
this low-quality  observation has  not been published.   The available
data indicate that the period of AB is 1.35\,years; the pair completed
nearly two revolutions since its first resolution in 2011.

{\it 02098$-$4052.} HIP 10096 is  SB2 according to GCS. The separation
corresponds  to a period  of $\sim$5\,yr.  

{\it 03046$-$5119.} HIP 14307 and 14313 form the 38\arcsec ~pair DUN~10
AB belonging to the FG-67 sample.  The component B is resolved here at
0\farcs19. The estimated period of Ba,Bb is $\sim$25\,yr. No motion is
seen in 2 months. The subsystem Ba,Bb is also manifested by asymmetric
line profiles (Tokovinin, in preparation). The A component has never
been observed at high angular resolution.

{\it 04386-0921.} HIP~21265  is an X-ray source and  an SB2
according  to GCS  (one  observation  only). It  is  resolved here  at
56\,mas, which corresponds to a period of $\sim$5\,yr.

{\it  04469-6036.}   First  resolution  of  HIP~22229  at  0\farcs043,
$\Delta  I=1.2$.  The  measure  is uncertain,  but  the elongation  is
confirmed with 2-ms exposure and  is not seen in other observed stars,
so it is not caused by  vibration.  This is a triple system containing
an eclipsing  binary AL~Dor  of Algol type  with an  eccentric orbit
\citep{Bulut2007}.  The  separation implies  an orbital period  of the
outer system on the order of 5\,yr.

{\it 05354$-$0450.}   This is HIP~26237, HD~37018,  HR~1892, 42~Ori, a
young  star in  Orion  which has  not  been observed  at high  angular
resolution so far, according to INT4.  We resolved the known binary AB
=  DA~4 and discovered  the spectacular  subsystem Aa,Ab  at 0\farcs16
(Figure~\ref{fig:ACF}).

{\it  06303$-$5252. }  HIP  30995 is an SB2 according  to GCS.   The new
companion  at  0\farcs2 with  $\Delta  I  =  4.1$~mag is  unlikely  to
correspond to the SB2, so the system is probably triple.

{\it 06497-7433.}  HIP~32735 has an acceleration detected in HIP2, but
no  RV  data.   The  resolved   0\farcs3  pair  implies  a  period  of
$\sim$80\,yr.   The  spectral  type   K0IV  given  in  SIMBAD  may  be
inaccurate  because  the  luminosity   and  color  of  the  components
correspond  to  main  sequence stars  with masses  of 1.0  and 0.6
${\cal M}_\odot$.

{\it 06499$-$2806.}  The 0\farcs2  pair HDS~947 had not been observed since
its discovery  by {\it  Hipparcos}.  Here it  is revealed as  a triple
system  (Figure~\ref{fig:ACF})  with  comparable  separations  between
components. HDS~947  probably corresponds to AB,  while the fainter
C component is new.

{\it 07038-4334.}   HIP~34052 =  HD~53680 = HIP~34065C  = GJ~264  is a
spectroscopic  and astrometric  binary  for which  \citet{Sahlmann2011}
give an  orbit with  $P=1688$\,d = 4.62\,yr and  an estimated semi-major axis of
0\farcs15.   There  is  also   an  astrometric  orbit  \citep{Mkr2008}  with
$P=4.11$\,yr that predicts $\theta =  327^\circ$ for the moment of our
first observation.  The  binary is resolved at 0\farcs23,  $\Delta I =
4.4$\,mag and shows some orbital  motion. The binary makes a quadruple
system together with components A and B (GJ 264.1 and 264) that form a
20\farcs5  pair at  185\arcsec  ~from  C.  The  B component  was  also
observed here and found unresolved.

{\it  07294-1500}  is  another   nearby  multiple  system.   The  main
component HIP~36395 is a visual binary with a known 728-yr orbit, also
measured here.  The C component (NLTT~17952) at 20\farcs4 is physical,
and yet another CPM component  F is found at 1072\arcsec ~\citep{LEP},
while  the WDS  components D  and E  are optical.   We observed  C and
resolved it at 0\farcs09.  The orbital period of Ca,Cb is on the order
of 6\,yr, estimated masses are  0.6 and 0.4 ${\cal M}_\odot$.  We also
targeted F and did not resolve it.  The magnitudes and colors of C and
F are quite similar.

{\it  07304+1352}  is  a  quadruple  system.   The  7\farcs7  pair  AB
(STF~1102) is HIP~36485, the CPM component D = HIP~36497 = HD~59450 is
located at 112\arcsec  ~from it, while the WDS components  C and E are
optical.   The physical  nature of  AD  is established  by common  PM,
distance,  and  RV.   D   is  a  known  SB1  with  $P=2708$\,d=7.4\,yr
\citep{Halb2012} and  an estimated semi-major axis of 93\,mas,  also an acceleration
binary.  We resolved the Da,Db pair at 0\farcs11, $\Delta I$=2.6.  The
minimum mass of Db derived from its SB orbit is 0.27 ${\cal M}_\odot$,
while  we estimate the  masses of  Da and  Db as  1.05 and  0.6 ${\cal
  M}_\odot$ from  their luminosity.   A previous non-resolution of  D is
reported in INT4; it was also unresolved with Robo-AO \citep{RAO}.

{\it 07312+0210,} HIP~36557 = HD~59688.  According to  observations
by D.~Latham  (2012, private  communication), this is  a spectroscopic
triple with an  inner period of 70\,d  (double-lined, also detected  by GCS,
mass ratio 0.7) and an outer  period of 2007\,d or 5.5\,yr. The outer system
is also detected by astrometric acceleration \citep{MK05}.  We resolve
it here at 0\farcs057, $\Delta I$=2.0, $\Delta y$=2.5 mag, and see the
orbital motion.   The estimated mass  of Ab is 0.88  ${\cal M}_\odot$.
The  semi-major axis  of the  70-d inner  binary Aa,Ab  is  7\,mas, so
accurate measurements of AB can detect the sub-motion to determine the
orientation of the inner orbit.

{\it 08021-1710.}  This is the high-PM M-dwarf LP~784-12 (HIP~39293) at
30\,pc from the Sun.  A new  distant component C was found at 1\farcs8
in addition to  the known pair HDS~1140 which  closed from 0\farcs4 in
1991.25 to  0\farcs33 now  (Figure~\ref{fig:ACF}). It is  confirmed as
physical by  its fixed position during  one year, the  quadrant of the
triple was determined in  run 5.

{\it  08447-2126.}  Like  the  previous object,  the  late-type
nearby binary  HDS~1260 was discovered by {\it Hipparcos}  and is expected to
move rapidly (HIP~42910, BD$-20^\circ$~2665).  It was targeted at SOAR
for the  first time.  To our surprise,  the object turned out  to be a
resolved triple,  with the secondary  being a 0\farcs16 pair  of equal
stars (Figure~\ref{fig:ACF}).  {\it Hipparcos} failed to recognize the
triple nature  of this  star. The estimated period  of BC is  15\,yr.  The
outer pair AB  has closed  from 0\farcs8 to  0\farcs5 and moved in
position   angle  since  its   discovery.   The   separations  between
components are  comparable, so this  triple system may  be interesting
dynamically.

{\it 09299$-$3629.} HIP 46572 is called a ``high proper motion star'' in
SIMBAD, although its PM and RV are actually quite moderate. The binary
AB  has moved   little  in the  24  years  since   its  resolution   by  {\it
  Hipparcos}. We discover the subsystem BC (Figure~\ref{fig:ACF})
with an estimated period of $\sim$100\,yr.

{\it 09586$-$2420. }  HIP 48906  is a double-lined binary according to
the GCS,  first   resolved  here  at   64\,mas.   The  period   should  be
$\sim$20\,yr.

{\it  10056-8405.}   HIP~49442 =  HD~88948  is  a  nearby dwarf  in  the
3\farcs9 visual  binary HJ~4310~AB. According to  GCS, the RV  of the
main  component A  varies by  3.5\,km~s$^{-1}$. Here  it is  resolved  into a
0\farcs18 Aa,Ab   pair with an estimated  orbital period  of  25\,yr.  No
astrometric acceleration was  detected, however.  The visual secondary
B was targeted separately and found unresolved.

{\it 10070-7129.} HIP~49546 is an astrometric binary of 1.5\,yr period
\citep{Goldin2006}  with variable  RV. The  period corresponds  to a
semi-major axis of  25\,mas. The star is resolved  here tentatively at
26\,mas (with  2-ms exposure).  This resolution is below the diffraction
limit and needs  confirmation. The measured position  angle of $346^\circ$
is close to $342^\circ$ predicted by the astrometric orbit.

{\it 10223-1032.}  HIP~50796 is  a single-lined and astrometric binary
according  to \citet{Torres2006},  with  period 570.98\,d  (1.56\,yr),
$K_1=20.76$\,km~s$^{-1}$, $e=0.611$.  The  {\it Hipparcos} parallax corrected
for  the  binary   motion  is  20.6$\pm$1.9\,mas.   The  spectroscopic
secondary  companion is  over-massive,  most likely  a  close pair  of
M-dwarfs. If so, the new  speckle companion at 1\farcs66 with a period
on  the order  of  500\,yr  makes the  system  quadruple. The  speckle
companion  might contribute  to the  IR  excess found  by Torres.  The
system is an X-ray source, and is possibly young.

{\it 10530+0458. } Latham (2012, private communication) identified HIP
~5212  =  HD~94292 as  SB2  with  a period  of  8.3\,yr  and a  highly
eccentric orbit. It is resolved here securely at 50\,mas.

{\it 12176+1427.   }  HIP~59933  has a variable  RV according  to GCS.
The   32\,mas  separation   corresponds  to   an  orbital   period  of
$\sim$2\,yr.  However, the separations in  the $y$ and $I$ filters are
somewhat discordant; further confirmation is needed.

{\it  12250-0414.} HIP~60574  is a  spectroscopic triple  with periods
14\,d  and  22\,yr  (Latham,  2012, private  communication),  also  an
acceleration  binary.   We  resolved   the  outer  pair  at  0\farcs22
separation,  matching  the spectroscopic  period.   The  lines of  the
visual  secondary Ab  could potentially  be detected  in  the spectrum  by
cross-correlation, leading eventually to a full 3-D orbit.

{\it  12528+1225. }  HIP  62933 (41  Vir) was  observed on  request by
F.~Fekel  who  studies  its  spectroscopic orbit.   Apparently  it  is
resolved for  the first  time. 

{\it  13132-0501.}   HIP~64499 has a  variable  RV,  with a  preliminary
spectroscopic   orbit   of  17\,yr   period   (Latham  2012,   private
communication).  It is resolved at 0\farcs1 and shows no motion in one
month.

{\it 13321-1115.}   No previous  indication of binarity  was available
for  HIP~66018, apart  from  the  CPM companion  B  at 84\arcsec.   We
discovered another  faint component  Ab at 0\farcs89,  $\Delta I$=4.6,
likely  to be  physical  (low background  density).   The B component 
($V=14.8$)  was targeted,  but  its  speckle signal  was  weak and  no
obvious  close  companions  to  B  were found.   

{\it  13344-5931.}  HIP~66230  has a  variable RV  in the  GCS  and an
astrometric acceleration. We resolved  it at 0\farcs1, $\Delta I$=2.5,
estimated  period $\sim$10\,yr.  The  pair moved  by $4^\circ$  in one
month.

{\it 13382-2341.}  HIP~66530  has a variable RV according  to the GCS.
It is  resolved at  0\farcs16 in the  $I$ band only,  estimated period
$\sim$20\,yr. This is a triple  system,  considering the CPM companion B
at 28\arcsec ~(LDS~4385~AB).

{\it 13401-6033.}   HIP~66676 (A) and HD~118735 (B,  G6V, $V=9.17$) at
77\arcsec ~share  common PM (although it is  small, 58\,mas~yr$^{-1}$) which,
together with photometry, indicates with high probability that it is a 
physical pair AB \citep{LEP}.  We targeted the secondary component B
and found it to be a  resolved triple.  The faint star C, at 0\farcs92
from    B,    is    itself    a    close    0\farcs16    pair    Ca,Cb
(Figure~\ref{fig:ACF}). Note that this is a region of the sky with very high
stellar density,  raising suspicion that  the Ca,Cb  pair might be a
random background object.  Re-observation  of the triple in 2015 
(to  be published)  shows, however,  that it  is physical  because the
center of C=(Ca,Cb) moves relative to B with a speed of 11\,mas~yr$^{-1}$ (or
3\,km~s$^{-1}$), compatible with  the expected orbital motion of  BC and much
less than the system's PM  of 58\,mas~yr$^{-1}$.  The estimated period of Ca,Cb
is $\sim$30 years, the period of BC is $\sim$300 years.

{\it  13495-2621.}   HIP~67458  is  a  double-lined  chromospherically
active  binary with orbital  period of  7.2\,d (Latham,  2012, private
communication). We found a  faint tertiary companion at 0\farcs73 with
an estimated orbital period on the  order of 100\,yr.  The speckle survey
of chromospherically  active stars  by \citet{Mason98} did  not detect
this tertiary, lacking the dynamic range of HRCam.

{\it 14014-3137. } HIP~68507 is an acceleration binary with a variable
RV resolved here at 0\farcs06. The period of Aa,Ab is on the order of
5\,yr.  There  is a  faint physical companion  B at 6\farcs7  found in
2MASS  \citep{ANDICAM}.    Another  visual  companion   at  14\arcsec,
~SEE~195, is optical, as revealed by its fast relative motion.

{\it  14382+1402.}  HIP~71572  is  an acceleration  binary without  RV
data. It is  found to be a  tight 90-mas pair with a  small $\Delta m$
and  an estimated  period  under  10\,yr  (some motion  is  seen  in  one
month). Very likely it can be studied as a double-lined SB.

{\it  14464-3346.  }   HIP~72235B  is located  at  9\arcsec ~from  the
primary  star   and  shares  its   proper  motion  (AB   =  HDS~2082).
Pre-discovery measurements of this {\em Hipparcos} pair were published
by \citet{Wycoff2006}.   The RV of  A may be variable  (Latham, 2012,
private communication).  The star B  turns out to be a 0\farcs4 binary
Ba,Bb with masses of 0.70 and 0.17 ${\cal M}_\odot$ estimated from the
luminosity.  Its period is on  the order of 100\,yr.  The whole system
could thus be quadruple.

{\it 15362-0623.} HIP~76400 is identified by the GCS as an SB2
with a mass ratio $q=0.93$, but there is no spectroscopic orbit
available.   We  resolved  the  0\farcs19 pair  with  $\Delta  I$=3.9,
indicating  a  mass  ratio  of  $\sim$0.5 and  an  orbital  period  of
$\sim$30\,yr.   Very likely  the resolved  binary does  not  match the
spectroscopic double-lined system. Considering  the CPM component B at
80\arcsec ~\citep{LEP}, this system could be a quadruple with a 3-tier
``3+1'' hierarchy.

{\it  15367-4208.} HIP~76435 is a G5V star from the FG-67 sample. Its
companion C (AC = FAL~78) at 13\farcs5 is physical, while the {\it
  Hipparcos} companion B at 4\farcs3  is not seen in
the 2MASS images and has not been confirmed otherwise. We targeted C
and resolved it into a close binary. Estimated masses of Ca and Cb are
0.70 and 0.66 ${\cal M}_\odot$, period $\sim$4\,yr.  

{\it  16142-5047.}  HIP~79576 has  a variable  RV (GCS).   The 79\,mas
separation  implies  an orbital  period  of  $\sim$5\,yr.   This is the
high-PM   star  LTT~6467  with   a  low   metallicity  [Fe/H]=$-$0.78.
\citet{Ivanov13} found no CPM companions.

{\it 16195-3054.}  HIP~79980 and  HIP~79979 form a 23\arcsec ~CPM pair
AB  where the  F6III  primary  is slightly  evolved,  while the  G1/G2
secondary is  closer to  the main sequence,  but still above  it.  The
{\it Hipparcos} parallax of B,  $-4.7$\,mas, is obviously wrong, so we
assume the parallax of the  primary, 20.7\,mas.  The RV(B) is variable
(GCS), and we resolve it into  a 40\,mas pair with an estimated period on
the order of 1\,yr. The pair moved by 10$^\circ$ in a month.  Binary
motion is the likely cause of the incorrect {\it Hipparcos} parallax.

{\it 16454-7150.}  HIP~82032 is located in a crowded field, so the new
faint Ab companion found here at  1\farcs3 could be optical. The star was
observed because  of its  suspected variable RV  (GCS), but  the newly
found companion,  even if physical, is  too distant to  explain this
variability. Another  visual companion B at  11\farcs5 (AB =  B~2392) is
optical, as evidenced by its fast  relative motion. The star is on the
HARPS exo-planet program.

{\it 16563-4040.} HIP~82876 is a  distant O7V star.  The 0\farcs26 pair
AB (HDS~2394)  was measured among other neglected  binaries.  We found
another  faint companion C  at 1\farcs46  (Figure~\ref{fig:ACF}).  The
star  has an extensive  literature, including  multiplicity  surveys with
speckle interferometry  and RV  \citep{Chini2012}. Owing to  the large
distance, no detectable orbital  motion is expected.  Indeed, the AB pair
was  measured with HRCam  in 2008.5 at  the same position as  it is
now. Those observations in the $y$ band did not detect the companion C
owing to a lower signal-to-noise ratio.

{\it 17054-3346.} The RV of HIP~83612 varies by 52~km~s$^{-1}$ (GCS). It is a
very close pair with an estimated period of $\sim$1.5\,yr, and the measure
near the  diffraction limit derived from the  elongated power spectrum
is tentative.   The {\it Hipparcos}  parallax is likely biased  by the
binary.  The visual component B  = HIP~83609 (AB = WNO~5) at 25\arcsec
~is optical.

{\it 17098-1031.}  HIP~83962 = HR~6375  has a variable RV according to
the GCS,  while Gorynya (2013, private  communication) detected double
lines.  It is resolved  tentatively at 33\,mas with $\Delta y=1.8$~mag
(the 5-ms exposure  makes it unlikely that the  asymmetry is caused by
vibrations).  The  separation corresponds to an orbital  period on the
order of  1\,yr, which  could bias the  {\it Hipparcos}  parallax.  
However, \citet{Eggleton08} consider the star as single. The
new pair was not resolved in 2014.3; presumably it became closer.

{\it 17264$-$4837.} HIP~85342 and HIP~85326 form a physical pair AB at
127\arcsec ~separation (common PM,  RV, and parallax). The B component
= HIP~85326  has a variable  RV and an astrometric  acceleration which
could hardly be  produced by the 1\arcsec ~speckle  companion Bb found
here, owing to  its long estimated period of  $\sim$300\,yr.  It seems
that B is triple and the whole system is quadruple.  This companion Bb
was not  detected in the  previous speckle observations because  it is
red: $\Delta I = 2.4$, $\Delta  y = 4$\,mag; its color matches a dwarf
star  at the  same  distance as  the  system.  However,  the field  is
crowded and the newly found companion could still be optical.

{\it  17266-3258.}   HIP~85360 is  an  acceleration and  spectroscopic
binary.   Its preliminary spectroscopic  period (Latham  2012, private
communication)  corresponds to  a semi-major  axis of  80\,mas.  The
star  is  chromospherically  active  and possibly  young.   The  faint
companion found here at 1\farcs16 is most likely optical, as the field
is extremely  crowded.  Re-observation within a year  will resolve its
status.

{\it 17341-0303.} HIP~85963  has a variable RV and  is an acceleration
binary.   The   91\,mas  separation  implies  an   orbital  period  of
$\sim$10\,yr; the estimated masses are 1.37 and 0.81 ${\cal M}_\odot$.
Despite extensive  literature (51 references  in SIMBAD), there  is no
published   spectroscopic   orbit,   while   several   high-resolution
spectroscopic studies addressed the abundance.

{\it  17342-1910.}  B~1863  is a  known  close binary  which has  been
unexpectedly  found to  be a  triple (Figure~\ref{fig:ACF}).   The new
distant  component C  is detectable  also in  the $y$  filter,  but we
measured only the  inner binary in $y$.  The star  was observed at the
Blanco telescope in 2008.5397, and the pair actually measured then was
AC,  at  133\fdg8, 0\farcs217,  $\Delta  y  =3.7$  (same as  now,  see
Table~\ref{tab:new}). The inner pair AB with  a smaller $\Delta
m$ was unresolved in 2008.5,  while it is clearly resolved now.  Owing
to the large  distance from the Sun, we expect only  a slow motion, so
even  the inner  pair observed  since 1929  may not  yet be  ready for
computing its first orbit.

{\it 17342-5454.}   HIP~85969 has a  variable RV according to  the GCS
and confirmed by  \citet{Jones2002}.  The 0\farcs55 separation implies
a  period on  the order  of  80\,yr.  The  star is  on the  exo-planet
program at the Anglo-Australian Telescope.

{\it 18243-0405.}   The neglected pair YSC~67  turned out to  be a new
triple  (Figure~\ref{fig:ACF}).   The outer  0\farcs35  binary AB  was
known  previously, now  we  detect elongation  that  implies an  inner
subsystem Aa,Ab.  We  compared with stars in the same  area of the sky
observed  before and after  to assure  that the  elongation is  not of
instrumental origin;  it is seen in  two filters. 

{\it 18267-3024.} HIP~90397 is an acceleration binary with a variable RV
(GCS), resolved  here at  69\,mas (estimated period  $\sim$4\,yr). The
star is targeted by exo-planet  programs.

{\it 18346-2734.} HIP~91075 was noted  as a double-lined binary by the
GCS  (one observation  only).   The separation  of  86\,mas implies  a
period of $\sim$10\,yr, while the double-lined observation matches the
moderate magnitude  difference (estimated masses 1.04  and 0.77 ${\cal
  M}_\odot$).  The star is an X-ray source, so a future combined orbit
could determine  masses for testing evolutionary models  of young stars.
The sky  around the object  is quite crowded; the  9\arcsec ~companion
I~1026 is optical (it moves too fast).

{\it 19206$-$0645.}   HIP 95068 is the neglected  {\it Hipparcos} binary
HDS~2735 AB, a distant K-type  giant. We did not resolve this 0\farcs1
binary, which remains unconfirmed,  but detected instead another faint
star at 1\arcsec. The stellar  background is crowded, the PM is small,
and the status of the new companion remains uncertain.

{\it  19209$-$3303.  }   HIP 95106  and  95110 form  the 13\farcs7  pair
HJ~5107 AB.  The  RV variability of A was suspected by the GCS, it is now
resolved as  a 0\farcs27 binary  with estimated period  of $\sim$35yr.
The component B was also  observed and found unresolved. It contains a
spectroscopic pair  (Tokovinin, in  preparation), the whole  system is
quadruple.

{\it  19221-2931.}   HIP~95203  is  another acceleration  binary  with
variable  RV  resolved  here.   The  relatively  large  separation  of
0\farcs77 corresponds  to a period  of $\sim$180\,yr. The  actual period
can be  as short as 60\,yr  if the pair  is seen now near  its maximum
separation (it would then have been closer at the time of the {\it Hipparcos}
mission).  Most likely, however, the faint visual companion found here
and the spectroscopic/acceleration pair make a triple system. There is
another companion  HIP~95164 at 435\arcsec.   The status of  this wide
pair (is it a real binary or  just two members of a moving group?)  is
not clear, but the association  of those stars leaves no doubt (common
RV, PM, and parallax).

{\it 19409-0152.}   HIP~96834 has a spectroscopic orbit  with a period
of 1\,yr and expected semi-major axis of 27\,mas (Latham 2012, private
communication).  We resolved this  pair, although the measurement near
the diffraction limit is uncertain. There are some unsolved questions,
however. Why, despite the small  magnitude difference $\Delta y = 1.2$
mag, were double  lines not seen?   Why, despite the  1-yr period, was the
{\it Hipparcos} parallax not strongly affected and the star appears
to be on the main sequence?

{\it 19453-6823.}  We resolved the  secondary component of HDS 2806 AB
into a  close pair Ba,Bb  (Figure~\ref{fig:ACF}).  This is a  K3 dwarf
within 50\,pc  from the Sun.  The  pair Ba,Bb should be  fast and turn
around in about 10\,yrs.

{\it  22259$-$7501.}  HIP  110712  and 110719  form the 20\farcs6  pair
DUN~238 AB. We observed both  components. The A component has a variable
RV according  to the GCS, but  not confirmed by \citet{Jones2002};  it was
unresolved. The  newly found pair Ba,Bb  has a period on  the order of
10\,yr.    Considering   the    distant   companion    C    found   by
\citet{Caballero2012}, the system contains at least 4 stars.

\section{New and improved orbits}
\label{sec:orb}

\begin{figure*}
\epsscale{1.1}
\plotone{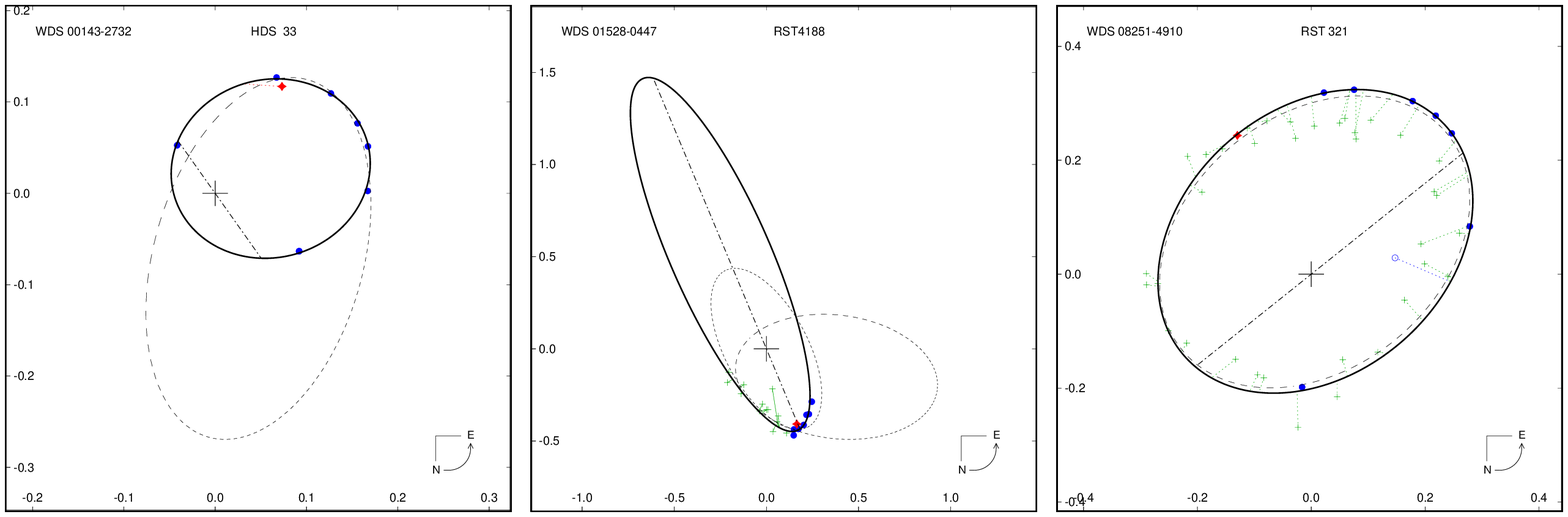}
\caption{\label{fig:rev}  Examples of revised  orbits. New  orbits are
  plotted in full line, previous orbit in dashed line. The position of
  the primary at the coordinate center is marked by a large cross, the
  line of  nodes is  traced by  a dash-dotted line.   The scale  is in
  arcseconds.  Interferometric (solid blue), {\it Hipparcos} (red) and
  micrometer (green crosses) measures are connected to their expected
  positions on the new orbit. A dotted ``O$-$C" line indicates a measure
  given zero weight in the orbit solution.
Left: a dramatic  orbit revision for {\it  01143$-$2732} (HDS~33, period
10.2\,yr).
Center: the long-period system {\it 01528$-$0447} (RST 4188, 619\,yr),
where  new  interferometric observations  caused  a substantial  orbit
revision; two more centuries of data are still needed to cover the extremity
of the ellipse and to constrain  period and semi-major axis.
Right:  a minor  revision  of  {\it  08251$-$4910}  (RST~321,  25.8\,yr)
demonstrating systematic errors of the historic micrometer measures.  }
\end{figure*}

\begin{figure*}
\epsscale{1.1}
\plotone{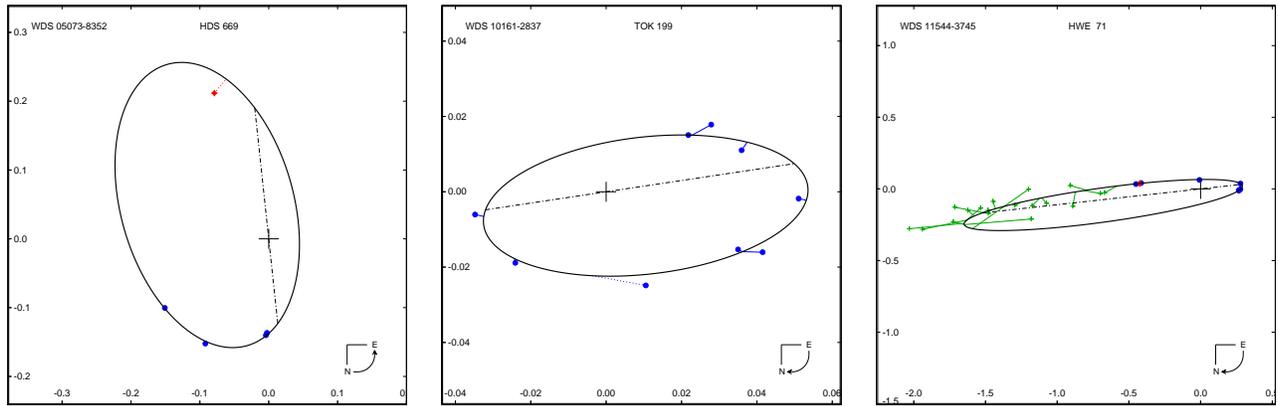}
\caption{\label{fig:first}  Examples of new  orbits (see the caption
  to Figure~\ref{fig:rev}). 
Left: the {\it Hipparcos} binary {\it 05073$-$8352} (HDS~699, period 26.5\,yr).
Center: ``fast'' spectroscopic binary {\it 10161$-$2837} (TOK 199, 2.5\,yr).
Right: visual binary {\it 11544$-$3745} (HWE 71, 249\,yr) that passed
recently through  periastron.   }
\end{figure*}

The  measurements reported  here  served to  improve  or compute  anew
orbits of 194 binary stars.  The orbital elements in standard notation
and their formal  errors are given in Table~5.   Its last column lists
the VB6  code to  a prior  orbit if it  exists (astrometric  orbits in
brackets). Errors  of the elements  (except provisional  orbits of
grade 5) are  given in the following line. For  the elements that were
fixed (from, for example, a spectroscopic orbit), an asterisk  replaces 
the error.  As discussed  in TMH14, orbit
improvements  range from  minor  ``cosmetic'' upgrades  to some  quite
drastic  revisions.   Blanks  in  the  last  column  indicate  the  43
first-time  orbit   solutions  (plus  three   with  prior  astrometric
orbits).  The grades  are assigned  according to  the  general grading
rules adopted in VB6 \citep{VB6}.  Most orbital elements
are derived using the USNO orbit code.

We do not publish  figures for all new orbits here, as they will be
available                           online                          in
VB6.\footnote{  \url{http://ad.usno.navy.mil/wds/orb6/wdsref.html}} 
Figure~\ref{fig:rev} illustrates orbit revisions ranging from dramatic
to    minor.    Three    first-time    orbits   are    presented    in
Figure~\ref{fig:first}.  Below we comment on some pairs.

{\it 04163$+$0710.}  WSI 97 is a single-lined nearby binary. Using the
radial velocities measured by D.~Latham (2012, private communication),
we computed a combined orbit  (the previous visual orbit reported by
\citet{RAO}  had  a  wrong   period).  The  inclination  is  close  to
$180^\circ$ and  had to be fixed  in order to match  the RV amplitude.
The complete orbit including RVs will be published later.

{\it   04506$+$1505.}   CHR   20  is   a  Hyades   binary   for  which
\citet{Griffin2012} published  an SB2 orbit. We combined  his RVs with
the speckle data, resulting in  a very accurate period. The combined
orbit corresponds to  a mass sum of 2.1 ${\cal  M}_\odot$ and an
orbital parallax of 22\,mas, in  good agreement with the HIP2 parallax
of $23.69 \pm 0.87$\,mas.

{\it    05245$-$0224.}   MCA~18~Aa,Ab   has    an   SB1    orbit   with
$P=9.44$\,yr. The orbit given here uses only the speckle data, however.

{\it  07490$-$2455.} The  period of  TOK 194  matches  the astrometric
orbit of  \citet{Gln2007}. The  mass sum  in this pair  composed of  a giant
primary and possibly an A-type  secondary is 5.9 ${\cal M}_\odot$. The
measure  on  2011.93 was  ignored  as  spurious  (it was  affected  by
vibrations).

{\it  07522$-$4035.} TOK~195  is the  bright star  a~Pup (HD  64440, HR
3080) known as a spectroscopic  binary.  However,  examination of the
RV  data  reveals  that   the  orbit  by  \citet{Parsons1983}  is  only
approximate. The binary  is difficult to measure, always  close to the
diffraction  limit  and  with  $\Delta   m  \sim  3$.  Instead  of  the
spectroscopic  period of 6.99\,yr,  our orbit  has $P=7.4$\,yr  and is
still preliminary. More RV coverage is obviously needed.

{\it 08391$-$5557.}  HU~1443~A,BC is  a triple system.  We provide the
first very  tentative orbit for the  outer binary, but  note its large
residuals  from the  recent measures  of AB.   Strictly  speaking, the
orbit should describe the motion of the center-of-gravity of BC around
A, rather  than the measures of  AB. Such refinement was  made for the
orbit of  A~2592~AB {\it (17156$-$0949)},  but it is not  warranted for
this preliminary orbit.

{\it  10161$-$2837.} TOK~199  is marked  as an  SB2 in  the  GCS, while
Latham  (2012, private  communication)  derived an  orbital period  of
916\,d, now independently confirmed by our orbit
(Figure~\ref{fig:first}). 

{\it 17157$-$0949.} This is the triple system HIP~84430. We computed the first
orbit of the secondary subsystem Ba,Bb which was discovered at SOAR in
2009 and has just completed one full revolution since.  Its separation
is always  close to  the diffraction limit.  Adopting a mass sum of
2.6 ${\cal M}_\odot$ for Ba,Bb, the resulting dynamical parallax is
7.6$\pm$1.5\,mas, while the HIP2 parallax is  4.9$\pm$0.9\,mas.
The latest orbit  of the outer pair A~2592~AB  published in TMH14 does
not account for  the fact that the speckle  measurements at SOAR refer
to A,Ba  and not to  AB. Here we  give a more accurate  solution that 
uses the positions of AB computed  from the measures of A,Ba under the
assumption that  Ba and Bb  have equal masses.  After  this correction
and  orbit   adjustment,  the  weighted  residuals   are  4.3\,mas  in
separation and 1\fdg3 in angle. Interestingly, there were a considerable
number  of speckle-interferometry measures  of this  pair obtained  in the
1980-s and 1990-s  at 4-m telescopes, but none  of them recognized the
subsystem Ba,Bb, despite its small  $\Delta m$.

Ignoring    the   multiplicity,    the    spectroscopic   survey    of
\citet{Guillout2009} determined a moderate  axial rotation $V \sin i =
10.8$\,km~s$^{-1}$  and  detected   the  lithium  line  of  52.8\,m\AA
~equivalent width  which, together with  the X-ray detection  by ROSAT
(RasTyc~1715-0948),  normally indicates youth.   These authors  do not
mention  this  star in  particular,  but  discuss  a group  of  active
lithium-rich giants  in their sample, to which  this system apparently
belongs. Even with the larger  dynamical parallax (instead of the HIP2
parallax),  all three  resolved  components of  HIP~84430 are  located
above the main sequence in the color-magnitude diagram.  This multiple
system is peculiar and merits further study.

\section{Summary}
\label{sec:concl}

We present here a large set of new speckle interferometry measurements
of close  binary stars, mostly with southern  declinations.  The total
number of  measurements made with  HRCam since 2008 now  exceeds 5000.
This unique data set is used  for calculation of 46 new orbits and for
improvement of  148 known orbits.   For comparison, the data  in TMH14
resulted in 13  new orbits and in the  improvement (sometimes drastic)
of  45 previously  known  orbits.  We  demonstrate  the good  internal
consistency of speckle astrometry  with HRCam by repeated measurements
of  relatively wide  binaries.  Typical  errors  are on  the order  of
3\,mas even for these wide pairs.

The high angular resolution and  dynamic range of HRCam give access to
close binaries never resolved before.  Some of those objects had prior
indication of  binarity from variable RV  or astrometric acceleration.
In such  cases, direct resolution allows us  to estimate statistically
orbital  periods   (which  are   typically  short)  and   to  evaluate
masses. This clarifies the statistics  of binary and multiple stars in
the solar  neighborhood.  We also  resolved a number of  components in
previously known nearby wide  binaries, converting them into triple or
higher-order hierarchies. 

A total of 56 newly-resolved pairs  are reported here, ten  of those being
inner or  outer subsystems in  visual binaries (Figure~\ref{fig:ACF}).
Most of  those subsystems  are totally unexpected.  Some of  the newly
resolved binaries  or subsystems are interesting  for various reasons,
such  as   being  young  (e.g.   X-ray   sources),  having  comparable
separations  and approaching  the dynamical  stability limit,  such as
HIP~9497 with periods of 138  and 13.9 years (TMH14), or being targets
of exo-planet programs.

\acknowledgments 

We thank the operators  of SOAR D.~Maturana, P.~Ugarte, S.~Pizarro, and
J.~Espinoza for efficient support  of our program. G.~Cecil has kindly
loaned us his  Luca-R detector which was used  for five nights instead
of our own broken camera.

R.A.M. acknowledges  support from the Chilean Centro  de Excelencia en
Astrof\'{\i}sica  y Tecnolog\'{\i}as Afines  (CATA) BASAL  PFB/06, and
the Project IC120009 Millennium Institute of Astrophysics (MAS) of the
Iniciativa  Cient\'{\i}fica Milenio  del Ministerio  de Econom\'{\i}a,
Fomento  y Turismo  de Chile.  R.A.M also  acknowledges  ESO/Chile for
hosting him during his sabbatical leave throughout 2014.

This work  used the  SIMBAD service operated  by Centre  des Donn\'ees
Stellaires  (Strasbourg, France),  bibliographic  references from  the
Astrophysics Data  System maintained  by SAO/NASA, and  the Washington
Double Star Catalog maintained at USNO.

{\it Facilities:}  \facility{SOAR}.

\clearpage


\begin{deluxetable*}{l l l  ccc  rc cc l r r l }                                                                                                                                
\tabletypesize{\scriptsize}   
\tablenum{2}                                                                                                                                                  

\tablecaption{Measurements of double stars at SOAR  (fragment)                                                                                                                              
\label{tab:double} }                                                                                                                                                            
\tablewidth{0pt}                                                                                                                                                                
\tablehead{                                                                                                                                                                     
\colhead{WDS} & \colhead{Discoverer} & \colhead{Other} & \colhead{Epoch} & \colhead{Filt} & \colhead{N} & \colhead{$\theta$} & \colhead{$\rho \sigma_{\theta}$} &               
\colhead{$\rho$} & \colhead{$\sigma \rho$} & \colhead{$\Delta m$} & \colhead{[O$-$C]$_{\theta}$} & \colhead{[O$-$C]$_{\rho}$} & \colhead{Reference} \\         
\colhead{(2000)} & \colhead{Designation} & \colhead{name} & +2000 & & & \colhead{(deg)} & (mas) & ($''$) & (mas) & (mag) & \colhead{(deg)} & \colhead{($''$)}                   
& \colhead{code$^*$}  }   
\startdata     
00026$-$0829 &	A 428 &	HIP 210 &	14.7632 &	I &	2 &	317.6 &	0.3 &	0.1197 &	0.9 &	0.4 : &	$-$1.0 &	0.003 &	This work  \\ 
            &	       &	           &	14.8535 &	I &	2 &	315.6 &	1.4 &	0.1165 &	0.8 &	0.3   &	$-$2.5 &	0.000 &	This work  \\ 
00028+0208 &	BU 281 AB &	HIP 223 &	14.8561 &	I &	2 &	160.3 &	0.2 &	1.5938 &	0.2 &	1.4 *  &  &   &   \\ 
00058$-$6833 &	HDS 4 &	HIP 488 &	14.7661 &	I &	2 &	229.3 &	4.8 &	0.1971 &	1.9 &	1.9    &  &   &   \\ 
00115$-$5545 &	HDS 25 &	HIP 927 &	14.7661 &	I &	3 &	75.3 &	0.7 &	0.1546 &	0.6 &	0.8    &  &   &   \\ 
00135$-$3650 &	HDS 32 &	HIP 1083 &	14.7662 &	I &	2 &	226.3 &	1.0 &	0.2565 &	0.6 &	0.8    &  &   &   \\ 
00143$-$2732 &	HDS 33 &	HIP 1144 &	14.8534 &	I &	2 &	55.5 &	0.3 &	0.1117 &	0.3 &	0.9   &	0.6 &	$-$0.002 &	This work  \\
\enddata 
\end{deluxetable*}

\begin{deluxetable*}{l l l   c c c  c c c c  }                                                     
\tabletypesize{\scriptsize}               
\tablenum{3}                                                      
\tablecaption{Unresolved stars  (fragment)                                                                 
\label{tab:single} }                                                                            
\tablewidth{0pt}                                                                                
\tablehead{                                                                                     
WDS (2000) & \colhead{Discoverer} & \colhead{Hipparcos} & \colhead{Epoch} & \colhead{Filter} &  
\colhead{N} & \colhead{$\rho_{\rm min}$} &    \multicolumn{2}{c}{5$\sigma$ Detection Limit} & $\Delta m$   \\             
$\alpha$, $\delta$ (J2000)  & \colhead{Designation} & \colhead{or other} & \colhead{+2000} & & &                           
& \colhead{$\Delta m (0\farcs15)$}  & \colhead{$\Delta m (1'')$}    & flag    \\             
 & \colhead{or other name}  & \colhead{name}  & & & & (arcsec) & \colhead{(mag)} &  \colhead{(mag)}  &     
}                                                                                               
\startdata     
00006$-$6641 &	GLI 289 &	HIP 55 &	14.7661 &	I &	2 &	0.039 &	1.95 &	3.07  \\ 
00052$-$6251 &	HIP 425 &	HIP 425 &	14.7661 &	I &	2 &	0.039 &	2.78 &	4.10  \\ 
00174$-$5131 &	HDS 40 &	HIP 1393 &	14.7662 &	I &	2 &	0.039 &	2.56 &	3.97  \\ 
00250$-$3042 &	HIP 1976 &	HIP 1976 &	14.8534 &	I &	2 &	0.039 &	3.42 &	4.82  \\ 
00301+0452 &	HIP 2358 &	HIP 2358 &	14.8561 &	I &	2 &	0.039 &	2.28 &	3.69  \\ 
00310$-$3138 &	HDS 69 &	HIP 2433 &	14.7662 &	I &	2 &	0.039 &	1.69 &	4.01  \\ 
00324+0657 &	MCA 1 Aa,Ab &	HIP 2548 &	14.7632 &	I &	2 &	0.039 &	3.01 &	4.25  
\enddata 
\end{deluxetable*}

\clearpage


\begin{deluxetable*}{l l l  cc  r r l lr }                                                                                                                                
\tabletypesize{\scriptsize}                                                                                                                                                     
\tablenum{4}
\tablecaption{Newly resolved   pairs
\label{tab:new} }                    
\tablewidth{0pt}  
\tablehead{
\colhead{WDS} & \colhead{Discoverer} & \colhead{Other} & \colhead{Epoch}  &  \colhead{Filt} &\colhead{$\theta$} &                
\colhead{$\rho$}  & \colhead{$\Delta m$} & \colhead{Sp.} &  \colhead{$\pi_{\rm HIP2}$}  \\         
\colhead{(2000)} & \colhead{Designation} & \colhead{name} & +2000 & &
\colhead{(deg)} &  ($''$) &  (mag) & type & (mas) }                       
\startdata 
01379$-$8259 &	TOK 426 &	HIP 7601 &	14.7661 &	I &	112.8 &	        0.0640 & 0.5  & G1V   & 36.5 \\ 
02098$-$4052 &	TOK 427 &	HIP 10096 &	14.7689 &	I &	218.3 &		0.0397 & 1.4  & F3V   & 14.1 \\ 
03046$-$5119 &	TOK 428 Ba,Bb &	HIP 14313 &	14.7635 &	I &	107.9 &		0.1911 & 0.5 :& G6V?  & 18.7 \\
04386$-$0921 &	TOK 387 &	HIP 21625 &	14.0457 &	I &    47.0 &		0.0557 & 0.6  & F8?   & 16.1 \\ 
04469$-$6036 &	TOK 388 &	HIP 22229 &	14.0456 &	I &    266.2 &		0.0431 &1.2   & F8V   & 16.1 \\
05354$-$0450 &	TOK 430 Aa,Ab &	HIP 26237 &	14.8567 &	I &	6.2 &	        0.1589 & 0.2  & B1V   & 3.7  \\
06303$-$5252 &	TOK 435 Aa,Ab &	HIP 30995 &	15.0288 &	I &	176.1&       	0.2134 &4.1 q & G4V   & 14.0 \\ 
06497$-$7433 &	TOK 389 &	HIP 32735 &	14.0597 &	I &	330.7 &		0.3343 &3.1   & K0IV  & 15.7 \\
\enddata
\end{deluxetable*}

\clearpage

\LongTables

\begin{deluxetable*}{l l rrr rrr r ccll}
\tabletypesize{\scriptsize}
\tablenum{5}
\tablewidth{0pt}
\tablecaption{Orbital Elements \label{tab:orb} }
\tablehead{
\colhead{WDS} & 
\colhead{Discoverer} &
\colhead{~~~P} & 
\colhead{~~T$_0$} &
\colhead{~~e} & 
\colhead{~~$\Omega$} &
\colhead{~a} & 
\colhead{~~i} &
\colhead{~~$\omega$} &
\colhead{Gr} &
\colhead{Orbit} \\
\colhead{(Figure)} &
\colhead{Designation} &
\colhead{~~~(yr)} & 
\colhead{~~(yr)} &
\colhead{ } & 
\colhead{~~($\circ$)} &
\colhead{~($''$)} & 
\colhead{~~($\circ$)} &
\colhead{~~($\circ$)} &
\colhead{ } & 
\colhead{Reference} 
}
\startdata
00026$-$0829 & A 428          &      414.95\phn &     2005.35\phn &      0.503 &     192.2 &      0.505\phn &      111.9 &       226.8 & 4 & Zul1984b   \\
             &                &  $\pm$20.96\phn &   $\pm$2.47\phn & $\pm$0.013 &  $\pm$2.0 & $\pm$0.021\phn &   $\pm$1.2 &    $\pm$4.2 &   &            \\
00143$-$2732 & HDS 33         &       10.22\phn &     2013.47\phn &      0.609 &     215.3 &      0.124\phn &       28.4 &        80.0 & 2 & Cve2012    \\
             &                &   $\pm$0.07\phn &   $\pm$0.04\phn & $\pm$0.009 &  $\pm$5.3 & $\pm$0.002\phn &   $\pm$2.5 &    $\pm$4.3 &   &            \\
00155$-$1608 & HEI 299        &        4.550    &     1995.363    &      0.360 &      60.8 &      0.306\phn &      145.6 &       344.5 & 2 & Hry1998    \\        
             &                &   $\pm$0.002    &   $\pm$0.016    & $\pm$0.005 &  $\pm$1.4 & $\pm$0.002\phn &   $\pm$1.3 &    $\pm$2.2 &   &            \\                
00315$-$6257 & I 260 CD       &       44.73\phn &     2011.62\phn &      0.810 &     241.6 &      0.495\phn &      124.7 &       112.2 & 3 & Msn2001c   \\
             &                &   $\pm$0.34\phn &   $\pm$0.01\phn & $\pm$0.001 &  $\pm$0.2 & $\pm$0.001\phn &   $\pm$0.3 &    $\pm$0.2 &   &            \\
\enddata
\end{deluxetable*}

\end{document}